\documentclass{aa}

\usepackage{graphicx}
\usepackage{txfonts}
\begin{document}

    \title{Hanle effect in the solar Ba\,{\sc ii} D2 line: a
    diagnostic tool for chromospheric weak magnetic fields}

%    \subtitle{ }
    \author{M. Faurobert\inst{1}, M. Derouich\inst{2}, V. Bommier\inst{3} and  J. Arnaud\inst{1}}

    \offprints{M. Faurobert}

    \institute{Universit\'e de Nice Sophia Antipolis, Fizeau Laboratory (CNRS/UMR 6525),
    Parc Valrose, F-06108 Nice, France\\
              \email{marianne.faurobert@unice.fr}
         \and
 Institut d'Astrophysique Spatiale, CNRS-Universit\'e Paris-Sud 11, 91405 Orsay Cedex, France             
             \and
             LERMA, Observatoire de Paris-Meudon, CNRS UMR 8112, 5 place Jules Janssen, 92195 Meudon Cedex, France \\
                          \email{veronique.bommier@obspm.fr}}
    \date{In press in Astronomy and Astrophysics}
    \titlerunning{Hanle effect in the Ba\,{\sc ii} D2 resonance line}
    \authorrunning{M. Faurobert}
\abstract{The physics of the solar chromosphere depends in a crucial
way on its magnetic structure. However there are presently very few
direct magnetic field diagnostics available for this region.}{Here
we investigate the diagnostic potential of the Hanle effect on the
Ba\,{\sc ii} D2 line resonance polarization for the determination of
weak chromospheric turbulent magnetic fields.}{The line formation is
described with a non-LTE polarized radiative transfer model taking
into account partial frequency redistribution with an equivalent
two-level atom approximation, in the presence of depolarizing
collisions and the Hanle effect.We investigate the line sensitivity
to temperature variations in the low chromosphere and to elastic
collision with hydrogen atoms. We compare center-to-limb variations
of the intensity and linear polarization profiles observed at THEMIS
in 2007 to our numerical results.}{We show that the line resonance
polarization is very strongly affected by partial frequency
redistribution effects both in the line central peak and in the
wings. Complete frequency redistribution cannot reproduce the
polarization observed in the line wings. The line is weakly
sensitive to temperature differences between warm and cold
components of the chromosphere. The effects of elastic collisions
with hydrogen atoms and of alignment transfer due to multi-level
coupling with the metastable $^2D_{5/2}$ levels have been studied in
a recent paper showing that they depolarize the $^2P_{3/2}$ level of
the line.  In the conditions where the line is formed we estimate
the amount of depolarization due to this mechanism as a factor of
0.7 to 0.65. If we first neglect this effect and determine the
turbulent magnetic field strength required to account for the
observed line polarization, we obtain values between 20 G and 30 G.
We show that this overestimates the magnetic strength by a factor
between 1.7 and 2. Applying these correction factors to our previous
estimates we find that the turbulent magnetic field strength is
between 10 G and 18 G.}{Because of its low sensitivity to
temperature variations, the solar Ba\,{\sc ii} D2 line appears as a
very good candidate for the diagnosis of weak magnetic fields in the
low chromosphere ($ z \ge 900$ km) by means of the Hanle effect.}

\keywords{line: profile --  Sun: magnetic fields -- techniques:
polarimetric-- Sun: chromosphere--  Radiative Transfer--
Scattering}

\maketitle

%_____________________________________________________________________________________________________________

\section{Introduction}
\label{intro} Magnetic fields play a crucial role in the physics of
the solar chromosphere (Innes  et al. 2005), however we lack direct
reliable diagnostic tools for the measurement of the field in this
region. The Zeeman effect on chromospheric lines is not easily
interpreted because most \textbf{of} them are broad and formed under
non-LTE conditions. Moreover, the field strength decreases from the
photosphere to the chromosphere and the physical conditions are
highly inhomogeneous so that the measurements of the Zeeman
polarization requires high spatial and spectral resolution. In this
context, the Hanle effect can provide a valuable diagnostic tool
because it is sensitive to weaker fields than the Zeeman effect and
it does not average out in the presence of unresolved mixed polarity
fields. The Hanle effect affects lines formed by scattering of
photons under anisotropic illumination by modifying their linear
resonance polarization; it is thus likely to occur in chromospheric
conditions.

Here we investigate the diagnostic potential of the resonance line
of ionized barium at 455.4 nm. Its linear polarization has been
recorded outside active regions by several authors using different
instruments (Stenflo and Keller 1997, Malherbe {\it et al.} 2006).
The observations that we show in this paper were obtained at THEMIS
in July 2007.

Theoretical works dedicated to the physical interpretation of these
observations are still in progress. A very comprehensive study of
the magnetic sensitivity of the Ba\,{\sc ii} D1 and D2 lines, fully
taking into account the hyperfine structure of the odd isotopes of
barium, has been published by Belluzzi {\it et al.} (2007). However,
this work, devoted to the investigation of the various physical
mechanisms which can play a role in the line polarization, does not
take into account radiative transfer effects. The line is modeled
under the last scattering approximation and the anisotropy of the
radiation field which gives rise to resonance polarization is chosen
ad-hoc. Here we adopt a complementary approach: we focus on the
radiative transfer effects and neglect the hyperfine structure of
the odd isotopes of Ba\,{\sc ii}. This will not allow us to recover
the well-known triple peak structure of the line polarization
profile, but we shall concentrate on the central polarization peak,
which is due to the even isotopes. This approach is justified by the
study of Belluzzi {\it et al.}, which has shown that the central
polarization peak is sensitive to the Hanle effect of weak turbulent
magnetic fields, whereas the two secondary peaks due the odd
isotopes are not. Moreover, in the line wings the linear
polarization degrees of the even and odd isotope components are
identical. In other words, the linear polarization in the line wings
may be computed ignoring the hyperfine structure of the odd
isotopes.

In the absence of hyperfine structure the ground level of the even
isotopes of Ba\,{\sc ii} is non polarizable by radiation anisotropy;
this is no longer true for the odd isotopes. This might affect the
line polarization at the wavelengths of the two secondary peaks due
to the odd isotopes, but not in the central peak nor in the line
wings. Furthermore, it is very likely that any ground level
polarization would be destroyed by elastic collisions with hydrogen
atoms, as is the case for the metastable $^2D_{5/2}$ levels (see
Derouich 2008).

In Sect. 2 we present the observations that we performed at THEMIS
together with some comments on data reduction. In Sect. 3 we explain
our non-LTE radiative transfer modeling of the line, taking into
account partial frequency redistribution and the Hanle effect due to
turbulent magnetic fields. We show that partial frequency
redistribution plays an important role in the resonance polarization
profiles obtained close to the solar limb. In Sect. 4 we investigate
the diagnostic potential of the line polarization. We first test its
sensitivity to temperature variations in the low chromosphere,
because the temperature may be quite inhomogeneous and
time-dependent in this layer (see Avrett 1995). We use two different
models of the quiet solar atmosphere, namely the standard average
FALC model (Fontenla, Avrett and Loeser 1993) and the FALX model
(Avrett 1995) derived from infrared measurements in the CO molecule,
which is significantly cooler in the low chromosphere. We also deal
with the effect of the elastic collisions which both broaden and
depolarize the line. We show that at the altitude where the line is
formed, the hydrogen density is low enough not to play a significant
role in the line broadening. The line depolarization by elastic
collisions with hydrogen atoms has been studied in detail by
Derouich (2008). This study has shown that in low density media the
depolarization of the $^2P_{3/2}$ level is not due to the relaxation
of the alignment but to alignment transfer between the $^2P_{3/2}$
level and the metastable $^2D_{5/2}$ levels. The line polarization
is thus overestimated if one neglects multi-level alignment
transfer.  As no fully consistent multi-level polarized transfer
code is presently available, we propose to compute the line
polarization with an equivalent two-level atom model and to take
into account multi-level alignment transfer effects by introducing a
correction factor to our results. We give this correction factor for
typical values of the hydrogen density in the low chromosphere above
$z=900$ km. Then, we show how the center-to-limb variations of the
linear polarization in the central peak of the line may be used to
investigate weak turbulent magnetic fields in the low chromosphere
of the Sun.

\section{Observations and data reduction}

\begin{figure}
\includegraphics[width=8.8cm]{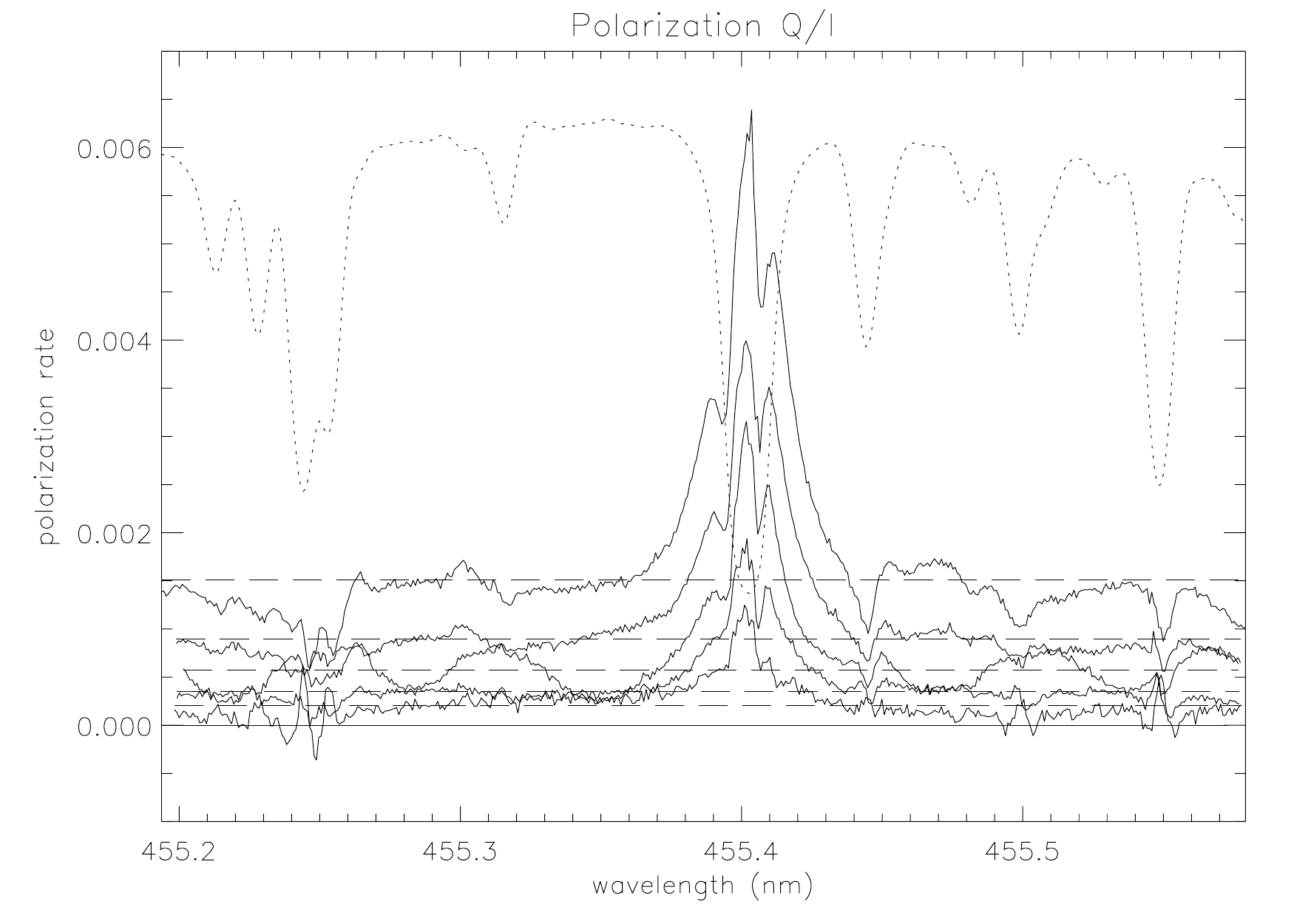}
\caption{Observation of the limb polarization of  Ba \,{\sc ii}
455.4 nm for a series of limb distances: from top to bottom, the
slit was positioned 4, 10, 20, 40 and 80 arcsec inside the limb, and
parallel to it. The data have been averaged along the slit.}
\label{fig-obs}
\end{figure}

The limb polarization of \ion{Ba}{ii} 4554 \AA\ was acquired with
the THEMIS telescope on 27 July 2007. The observation was performed
at the North Pole, and the slit was placed parallel to the limb, at
a series of limb distances (4, 10, 20, 40 and 80 arcsec). For the
short limb distances (4 and 10 arcsec), the limb distance was
determined in the slit-jaw images. For the larger limb distances,
the limb distance was manually allocated by using the pursuit tool
of the telescope. The THEMIS telescope was operating in the MTR-grid
mode, and the data were averaged along the slit. The polarization
was measured by applying the beam-exchange technique, and the data
reduction was done following Bommier \& Rayrole
(\cite{Bommier-Rayrole-02}) and Bommier \& Molodij
(\cite{Bommier-Molodij-02}), except for the fact that the THEMIS
polarimeter was later modified, now allowing beam-exchange for all
the 3 polarization Stokes parameters individually. No significant
signal was detected on Stokes $U$ and $V$. The observation results
for Stokes $Q$ are reported in Fig. \ref{fig-obs}. The continuum
polarization was adjusted to the theoretical value given by Fluri \&
Stenflo (\cite{Fluri-Stenflo-99}).

\section{Non-LTE line modeling}

In order to compute the line intensity and polarization profiles
both without and with the Hanle effect we use the same procedure as
the one described by Faurobert-Scholl (1992). In a first step we
solve the statistical equilibrium equations for the populations of
the atomic levels, coupled to the line and continuum transfer
equations, neglecting the polarization. This provides us with the
optical depth in the line of interest and with multi-level coupling
terms in its source function. In a second step, we compute the line
intensity and polarization with an equivalent two-level approach.
This two-step method relies on the fact that the line linear
polarization is weak enough not to play a role in the statistical
equilibrium of the atomic levels and on the hypothesis that the
alignment of the $^2P_{3/2}$ upper level is only due to optical
pumping from the unpolarized ground level, partially destroyed by
elastic collisions, but not affected by radiative coupling with
other atomic levels. However in the case of the Ba\,{\sc ii} D2 line
we have to take into account depolarizing effects due to alignment
transfer between the $^2P_{3/2}$ level and the metastable
$^2D_{5/2}$; this is done by introducing a correction factor on the
line polarization obtained from our equivalent two-level modeling.

The barium atom has odd and even isotopes with slightly different
atomic structures. The 134, 136 and 138 even isotopes (82.1\% of the
total number density) have no nuclear spin, i.e. no hyperfine
structure, whereas the 135 and 137 odd isotopes have a nuclear spin
$I=3/2$; the hyperfine structure of their fundamental level is
detectable (Rutten 1978), but we do not consider it here, as we deal
only with the central polarization peak due to the even isotopes.
Figure 2 shows the atomic model that we use, together with the
allowed radiative transitions.

\subsection{Polarization radiative transfer equation and
redistribution matrix in the presence of a weak magnetic field} The
polarized radiative transfer equation that we solve for the 455.4 nm
resonance line is written
\begin{equation}
\mu {\partial {\bf I}(\nu, {\bf n},z)\over\partial\tau_\nu}={\bf
I}(\nu, {\bf n},z)-{\bf S}(\nu, {\bf n},z),
\end{equation}
where $\mu$ is the cosine of the heliocentric angle of the line of
sight, ${\bf I}$ is the 2-component vector $(I,Q)^\dag$ of the
radiation field at frequency $\nu$, in the propagation direction
${\bf n}$ and at depth z in the atmosphere, $\tau_\nu$ is the
monochromatic optical depth at z. As usual $I$ is the specific
intensity of the radiation field and $Q$ is the Stokes parameter for
the linear polarization, defined by $Q= I_r-I_l$, where $I_r$
denotes the intensity along the radial direction and $I_l$ the
intensity along the direction parallel to the solar limb. The Stokes
parameters $U$ and $V$ vanish in the absence of a magnetic field and
in the presence of a mixed polarity field. The vector ${\bf S}$ is
the 2-component source function, given by
\begin{equation}
{\bf S}(\nu, {\bf n},z)={{{\bf j_c}({\bf n},z)+{\bf j_l}(\nu, {\bf
n},z)}\over{k_c(z)+k_l}(\nu, z)},
\end{equation}
where ${\bf j_c}$ and ${\bf j_l}$ denote the emissivity in the
continuum and in the line, respectively (notice that we take into
account the continuum polarization due to Thomson and Rayleigh
scattering), $k_c$ and $k_l=k_l^0(z)\phi(\nu,z)$ are the absorption
coefficients in the continuum and in the line respectively, $\phi$
is the line absorption profile. The line emissivity is detailed in
Faurobert-Scholl (1992), in a two-level atom formalism. It is
composed of 3 terms: a thermal emissivity, a term due to multi-level
coupling, and a scattering term which gives rise to linear
polarization in the line. The scattering term is written
\begin{equation}
{\bf j_{sc}}=k_l^0(z)\int_0^\infty d\nu'\int {d\Omega'\over
4\pi}\hat R(\nu,{\bf n}, \nu',{\bf n'},z){\bf I}(\nu',{\bf n'},z),
\end{equation}
where the matrix $\hat R$ is the redistribution matrix of the
polarized radiation field due to scattering processes. It describes
the correlation between frequency, direction of propagation and
state of polarization between incident and scattered photons. Its
analytical expression is given by Bommier (1997a, b), in the
presence of collisions and of a weak magnetic field. Here we use the
angle averaged form of the redistribution matrix, which corresponds
to the approximation level III of Bommier (1997b) (also see Fluri et
al. 2003 for numerical results). Bommier showed that the
redistribution matrix is the sum of several terms, corresponding to
the $R_{II}$ and $R_{III}$ types of scattering mechanisms, i.e.
respectively, coherent scattering or complete frequency
redistribution in the rest frame of the atom. The branching ratios
between those terms depend on the radiative and collisional
de-excitation rates, and they differ according to the frequency
domains of the incident and scattered photons, so that the
analytical expression of the redistribution matrix is quite
intricate. However, in the case of the Ba\,{\sc ii} D2 line, the
situation is simpler because the line is formed in dilute regions of
the low chromosphere where the radiative de-excitation rate is much
higher than the collision rates. This is illustrated in Fig. 2 which
shows the depth-dependence of the relevant branching ratios,
\begin{eqnarray}
\alpha & = & {\Gamma_R\over \Gamma_R + \Gamma_I +\Gamma_C},\cr
\beta^{(0)} & = &{\Gamma_R\over \Gamma_R + \Gamma_I },\cr
\beta^{(2)} & = &{\Gamma_R\over \Gamma_R + \Gamma_I + D^{(2)}},
\label{branch-ratio}
\end{eqnarray}
where $\Gamma_R$ is the radiative de-excitation rate, $\Gamma_C$ the
elastic collision rate and $D^{(2)}$ is the depolarizing collision
rate due to collisions with hydrogen atoms, $\Gamma_I$ is the
inelastic collision rate due to collisions with electrons. For the
Ba\,{\sc ii} D2 line, $\Gamma_R = 1.59\, 10^8$ s$^{-1}$, and
$\Gamma_C$ is derived from the collision cross-sections with
hydrogen atoms given by Barklem \& O'Mara (1998); the depolarizing
collision \textbf{rate is }obtained from a well tested
semi-classical method (Derouich, Sahal-Br\'echot and Barklem 2004),
\textbf{it} may be represented by the analytical expression
\begin{eqnarray}
D^{(2)} & = & 6.82\, 10^{-9}n_H(T/5000)^{0.40} \cr
 & + & 7.44\,
10^{-9}n_H(T/5000)^{0.38}\times (1/2)^{3/2}{\rm exp} (\Delta E/kT),
\end{eqnarray}
where $n_H$ is the neutral hydrogen density and $\Delta E$ is the
energy difference between the two fine structure levels $^2P_{1/2}$
and $^2P_{3/2}$.

We see in Fig. \ref{fig2} that at the altitudes where the line is
formed, i.e. above $z$= 700 km, the 3 branching ratios are close to
one. In that case, one can check that the $R_{III}$ type of
scattering gives zero polarization in the line wings and that the
$R_{II}$ one leads to the Hanle effect in the line core and Rayleigh
scattering in the line wings. This result was already obtained from
heuristic arguments by Stenflo (1994) who generalized the work of
Domke \& Hubeny (1988) to the magnetic case.

We recall that in the presence of an isotropic turbulent magnetic
field with a single value, the Hanle phase matrix reduces to
\begin{eqnarray}
{\hat P_H}({\bf n},{\bf n'},{\bf B})& = &{\hat P^{(0)}}({\bf n},{\bf
n'}) \cr & +& W_2[1-0.4 ({\Gamma_H^2\over
1+\Gamma_H^2}+{4\Gamma_H^2\over 1+4\Gamma_H^2}) ]{\hat P^{(2)}}({\bf
n},{\bf n'}), \label{hanlemat}
\end{eqnarray}
the so-called Hanle parameter is given by $\Gamma_H= 0.88g_J
B/(\Gamma_R + D^{(2)})$, where $B$ is in Gauss and $\Gamma_R$ and
$D^{(2)}$ are in units of $10^7$ s$^{-1}$. The polarizability
coefficient $W_2$ depends on the quantum numbers $J$ and $J'$ of the
line atomic levels, for the Ba\,{\sc ii} D2 line, $W_2 =0.5$. The
(2x2) matrices ${\hat P^{(0)}}$ and ${\hat P^{(2)}}$ are given by
\begin{equation}
{\hat P^{(0)}}= \pmatrix{1&0\cr0&0},
\end{equation}
and
\begin{equation}
{\hat P^{(2)}}= {3\over 8}\pmatrix{{1\over
3}(1-3\mu^2)(1-3\mu'^2)&(1-3\mu^2)(1-\mu'^2)\cr
 (1-\mu^2)((1-3\mu'^2)&3((1-\mu^2)((1-\mu'^2)},
\end{equation}
where $\mu$ and $\mu'$ denote the cosine of the colatitudes of the
scattered and incident radiation field, respectively. The Rayleigh
phase matrix has a similar expression as in Eq. (\ref{hanlemat}),
for $\Gamma_H=0$ and $W_2=1$.

\begin{figure}
   \centering
   \includegraphics[width=5cm]{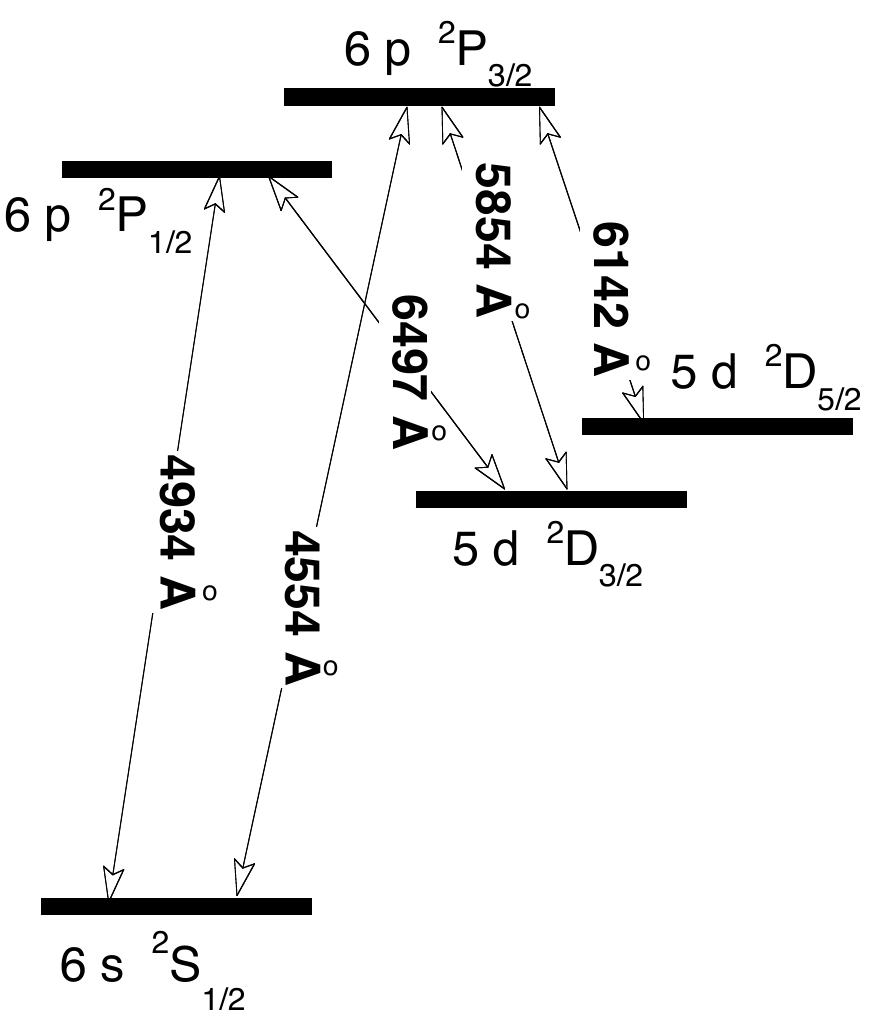}
   \caption{Ba\,{\sc ii} simplified atomic model, the Ba\,{\sc iii} ionization
continuum is taken into account but not shown in the Fig.}
\label{fig1}
    \end{figure}

%\begin{figure}[ht]
%\resizebox{1.0\hsize}{!}{\includegraphics[trim = 0mm 0mm 0mm
%0mm]{6302_v0_image_sp.ps}}
%  \caption{A .}
%  \label{figure1}
%\end{figure}
%%%%%%%%%%%%%%%%%%%%%%%%%%%%%%%%%%%%%%%%%%%%%%%%%%%%%%%%%%%%fig1

\begin{figure}
   \centering
   \includegraphics[width=8.8cm]{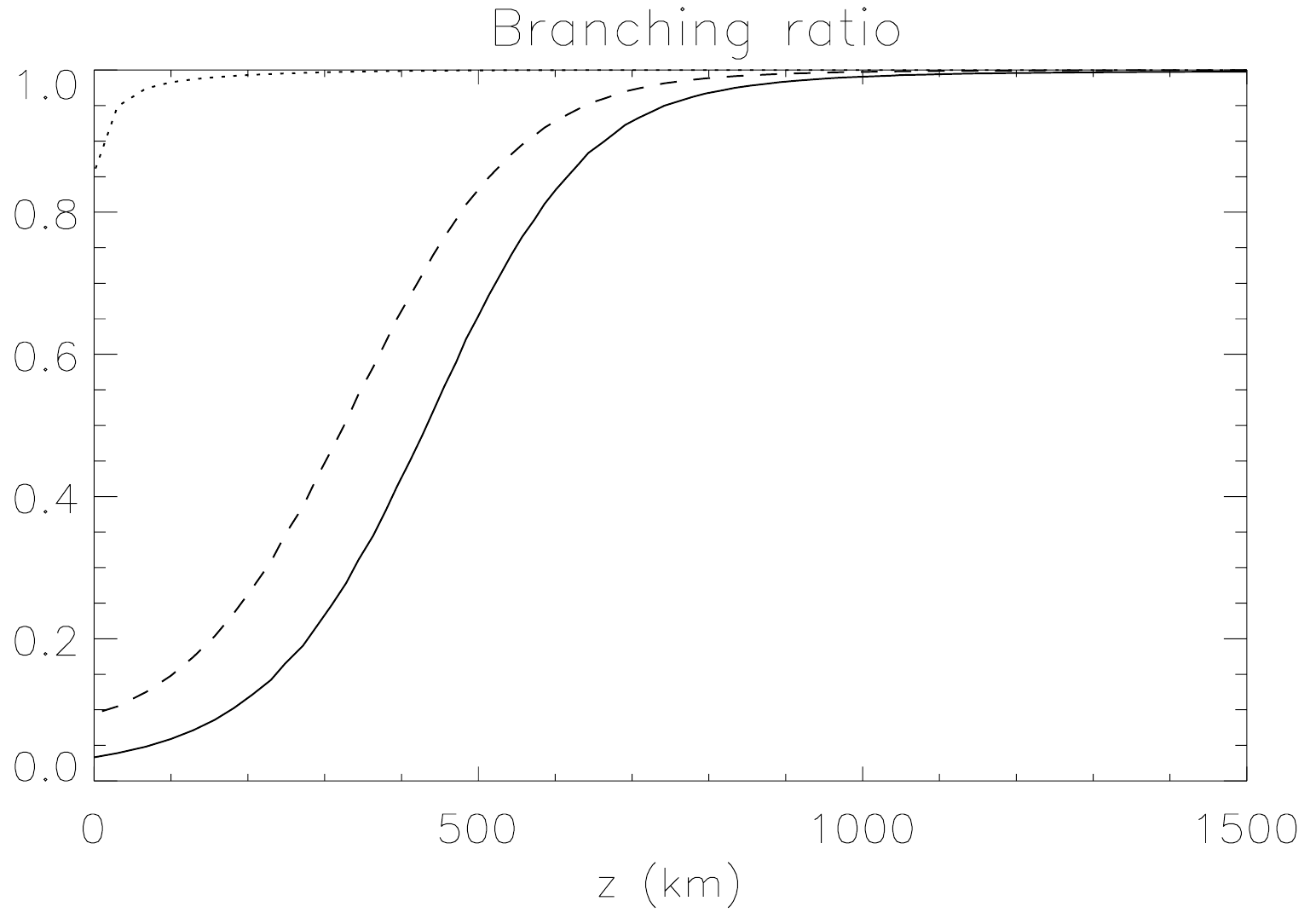}
   \caption{Depth-dependence of the branching ratios $\alpha$ (full line),
 $\beta^{(0)}$ (dotted line) and $\beta^{(2)}$ (dashed line), defined in Eq. (\ref{branch-ratio}).}
 \label{fig2}
    \end{figure}

\subsection{Partial frequency redistribution effects}
\begin{figure}
   \centering
   \includegraphics[width=8.8cm]{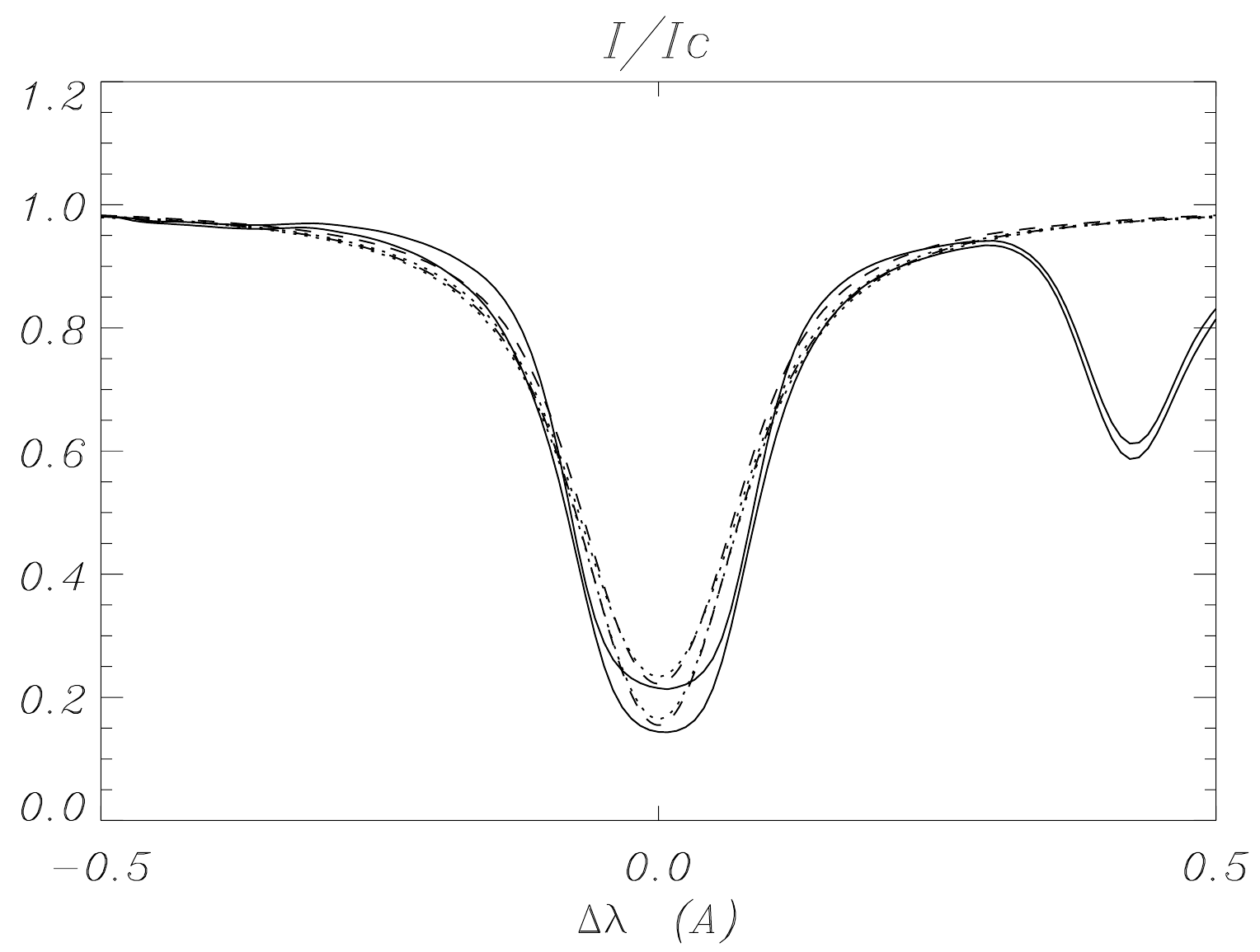}
   \caption{Partial frequency redistribution effects.
Intensity profiles for 2 values of the limb distance, d= 4" and 40".
{\it Full lines}: observed profiles, {\it dotted lines}: calculated
profiles with complete frequency redistribution, {\it dashed lines}:
calculated profiles with partial frequency redistribution.}
\label{partialI}
    \end{figure}

\begin{figure}
   \centering
   \includegraphics[width=8.8cm]{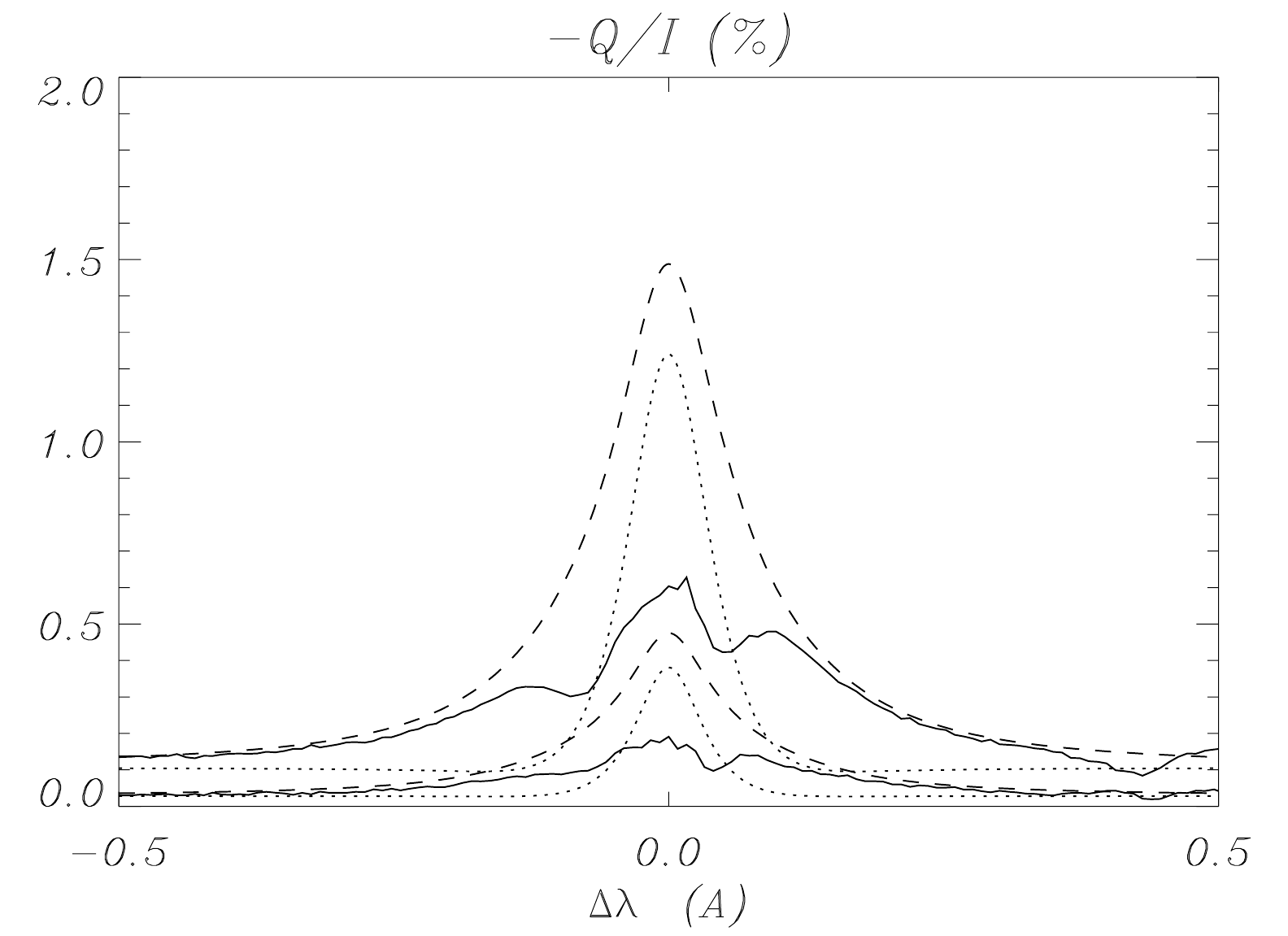}
   \caption{Partial frequency redistribution effects.
Linear polarization profiles $ -Q/I$ for 2 values of the limb
distance, d= 4" and 40". {\it Full lines}: observed profiles, {\it
dotted lines}: profiles calculated with complete frequency
redistribution, {\it dashed lines}: profiles calculated with partial
frequency redistribution.} \label{partialQ}
    \end{figure}

In Figs. \ref{partialI} and  \ref{partialQ} we compare the observed
intensity and polarization profiles to the results of our non-LTE
radiative transfer modeling both with partial frequency
redistribution (PFR) and with complete frequency redistribution
(CFR). In both cases the magnetic field is set to zero and the
standard FALC model of the quiet solar atmosphere is used. In order
to fit the observed intensity profiles one has to take into account
the effect of macroturbulent velocity fields in the solar
atmosphere. This amounts to a smearing of the profiles by a
normalized Gaussian function with a width parameter. We assume
typical values of the order of 3 km/s for the quadratic mean value
of the macroturbulent velocities.

The intensity profiles, shown in Fig.\ref{partialI}, are weakly
sensitive to partial frequency redistribution effects (see
Uitenbroek \& Bruls 1992 for a detailed analysis). We notice that
the line width is not well reproduced by our model, the reason is
that we have neglected the hyperfine splitting of the lower level of
the odd isotopes, which actually broadens the line profile. Fig.
\ref{partialQ} shows that the line polarization profiles are
strongly affected by partial frequency redistribution effects, both
in the line core and in the wings. As expected, we do not reproduce
the triple peak structure of the polarization peak either with PFR
or with CFR. The computed line polarization is significantly smaller
with CFR than with PFR. Complete frequency redistribution leads to
zero polarization in the line wings, in contradiction to the
observations, whereas the PFR model reproduces quite well the
observed polarization profiles in the wings.

\section{Investigation of the line polarization diagnostic potential }
\subsection{Sensitivity to temperature variations}
\begin{figure}
   \centering
   \includegraphics[width=8.8cm]{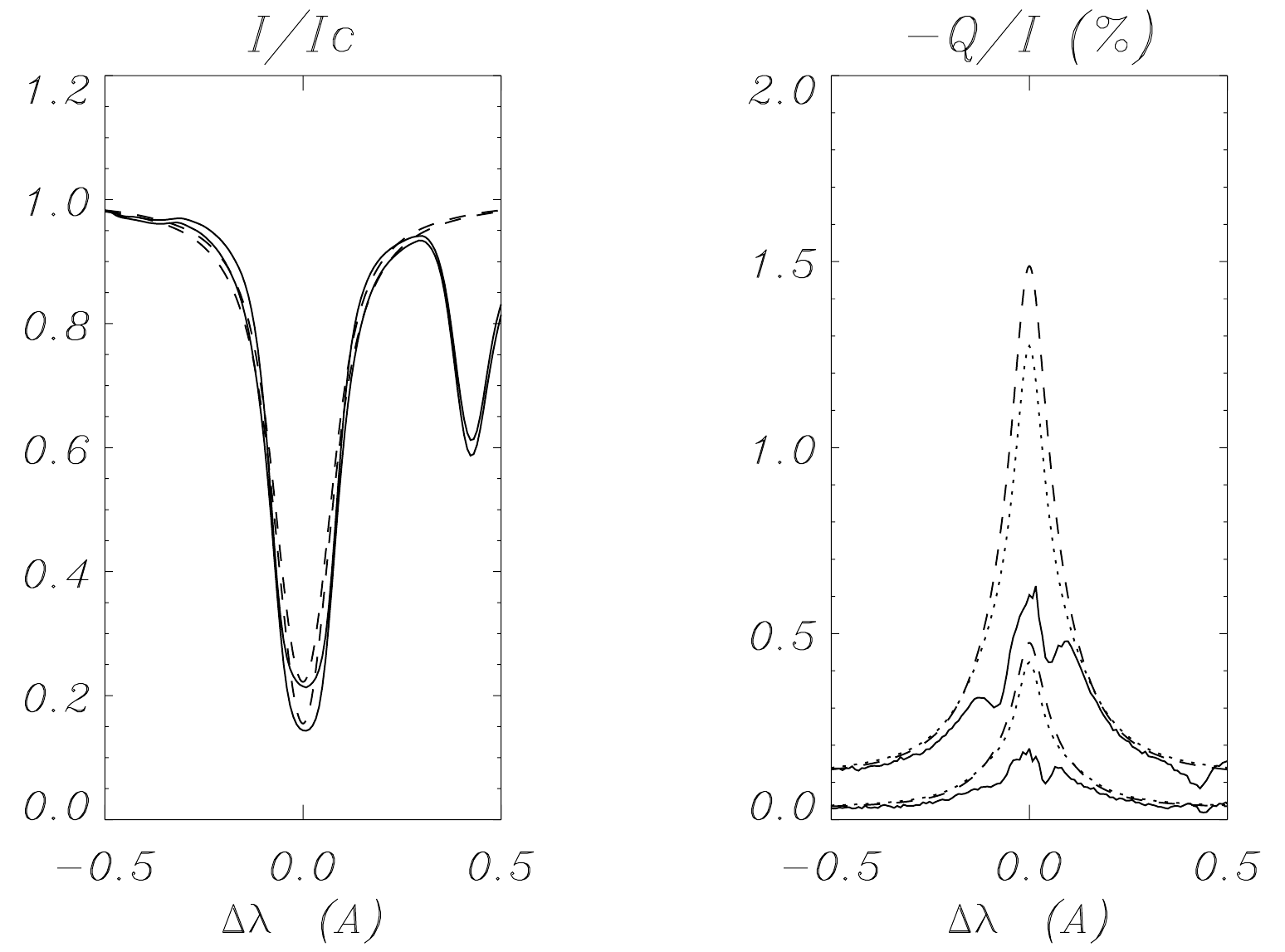}
   \caption{Intensity and
linear polarization profiles for 2 values of the limb distance, d=
4" and 40". {\it Full lines}: observed profiles, {\it dotted lines}:
profiles calculated with the FALX model, {\it dashed lines}:
profiles calculated with the FALC model.} \label{models}
    \end{figure}

The low solar chromosphere is an inhomogeneous medium, with cold and
hot components, varying with time (see Avrett 1995, Holzreuter et
al. 2006). In order to test the sensitivity of the line to
temperature variations we compare the intensity and polarization
profiles obtained for two models of the quiet solar atmosphere, i.e.
the standard average FALC model and the FALX cold model. As shown in
Fig. (\ref{models}), the intensity profiles obtained for the two
models are almost indistinguishable, the reason is probably that
non-LTE effects tend to decouple the line source function from the
Planck function. However the amplitude of the linear polarization
peak, which is related to the anisotropy of the radiation field,
shows a higher sensitivity to the atmospheric model. The colder
model leads to a slightly smaller value for the maximum of the peak
at line center, the effect becomes larger when ones approaches the
solar limb and at the limb distance $d=4"$ the linear polarization
peak reaches about 1.5\% with the FALC model and about 1.3\% with
FALX. We shall see in the following that these differences do not
play a significant role in the diagnosis of microturbulent magnetic
fields by means of their Hanle effect in the line.

\subsection{Sensitivity to elastic collisions}
\begin{figure}
   \centering
   \includegraphics[width=8.8cm]{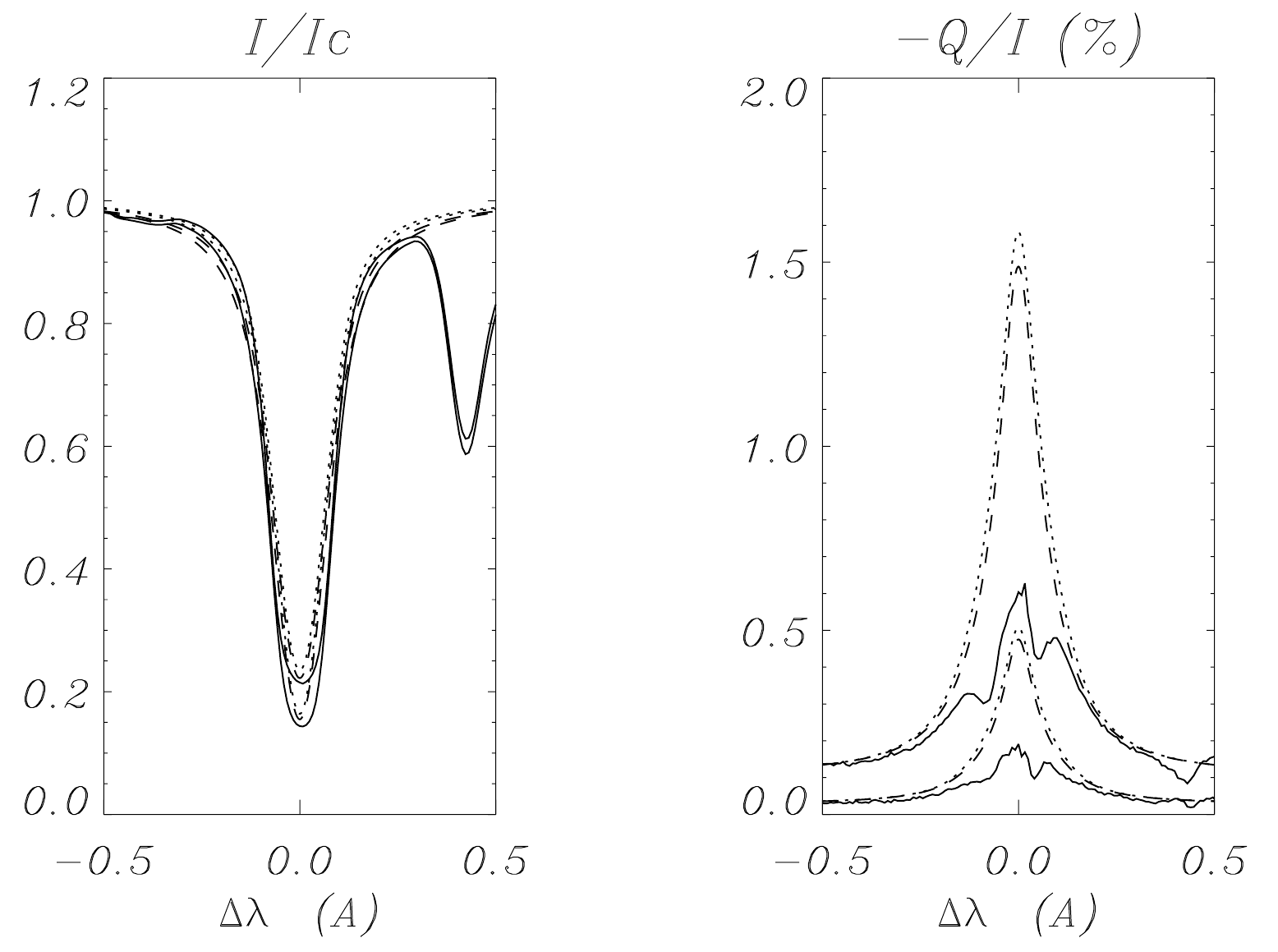}
   \caption{
Left panel: Intensity profiles for 2 values of the limb distance, d=
4" and 40". Right panel: linear polarization for the same values of
the limb distance. {\it Full lines}: observed profiles, {\it dotted
lines}: profiles calculated with the standard BOM value for the
elastic collision rate  $\Gamma_C= 3. \gamma_{VW}$, {\it dashed
lines}: profiles calculated for $\Gamma_C= 2. \gamma_{VW}$.}
\label{elasticrate}
    \end{figure}

The collisional cross-section obtained by Barklem \& O' Mara (BOM)
is 3 times larger than the classical Lindholm theory with the Van
der Waals approximation; they estimate the relative accuracy of
their results at about 10\%. We have tested the effect on the line
intensity and polarization of varying the elastic collision rate,
namely we used the two values $\Gamma_C= 3. \gamma_{VW}$ (BOM
standard result) and $\Gamma_C= 2. \gamma_{VW}$, where $\gamma_{VW}$
denotes the Van der Waals value. In both cases the calculations were
done with PFR and for the FALC model, the results are shown in Fig.
(\ref{elasticrate}). We see that there are very few differences
between the 2 cases, both in the line intensity and polarization
profiles. The line polarization is slightly smaller for the model
with the largest collision rate, which is to be expected, but the
effect is weak because, as explained previously, the line is formed
in a low density medium where radiative broadening dominates over
collisional broadening. We notice that in the line wings, which are
not sensitive to the Hanle effect, the polarization obtained with
the BOM value for the elastic collision rate is in slightly better
agreement with the observed profile.

As far as depolarizing collisions are concerned, recently Derouich
(2008) has shown that in the low chromosphere, where the $D^{(2)}$
depolarizing collision rate is much smaller than the line radiative
transition rate, the depolarization of the $^2P_{3/2}$ level is
mainly due to alignment transfer with the $D_{3/2,5/2}$ metastable
levels of Ba\,{\sc ii}. The corresponding depolarization increases
when the hydrogen density increases, because the alignments of the
metastable $^2D_{3/2,5/2}$ levels are vulnerable to collisions. In
order to estimate this effect we solved the statistical equilibrium
equations for the five BaII levels shown in Fig. 1 including
collisions, radiation and magnetic field effects as in Derouich
(2008) i.e. with a single scattering approximation, and we compared
with the results of the equivalent two-level model, for 3 values of
the hydrogen density. The results are given in the second column of
Table 1, where $\Delta p/p_{max}$ is the relative depolarization due
to alignment transfer and $p_{max}$ denotes the resonance
polarization obtained with an equivalent two-level model. We see
that the relative depolarization varies between 30\% and 40\%.

A physical interpretation of the effect of multi-level coupling on
the polarization of the Ba\,{\sc ii} D2 line is given here. By
solving the statistical equilibrium equations for the five BaII
levels shown in Fig. 1 we can estimate the relative importance of
the three absorption mechanisms responsible for the atomic
polarization of the $P_{3/2}$ upper level. We find that 69\% of
absorptions take place from the fundamental $S_{1/2}$ level, 3\%
from the $D_{3/2}$ metastable level and 28\% from the $D_{5/2}$ one.
The polarisability coefficient of those transitions are $W_2=0.5$,
0.32 and 0.02, respectively. The $D_{5/2} \to P_{3/2}$ transition
has a low polarisability and it is responsible for an important
fraction of the $P_{3/2}$ population. This leads to a decrease of
the polarization as compared to the results of the equivalent
two-level atom model where the only possible polarizing mechanism is
the absorption of radiation from the fundamental level.

\begin{table}
\caption{Depolarizing effect of multi-level alignment transfer on
the Ba\,{\sc ii} D2 linear polarization and correction factor $f_c$
on the magnetic field obtained with an equivalent two-level atom.}
\centering
\begin{tabular}{c c  c }
\hline \\
$n_H$ (cm$^{-3}$) & ${\Delta p\over p_{max}}$ &
$f_c$\\

\\
\hline\\
0. & 0.30&0.60\\
$3\, 10^{13}$&0.35&0.51\\
$5\, 10^{13}$&0.40&0.44\\
\hline
\end{tabular}
\label{table1}
\end{table}

\subsection{Hanle effect} 
\begin{figure}
   \centering
   \includegraphics[width=8.8cm]{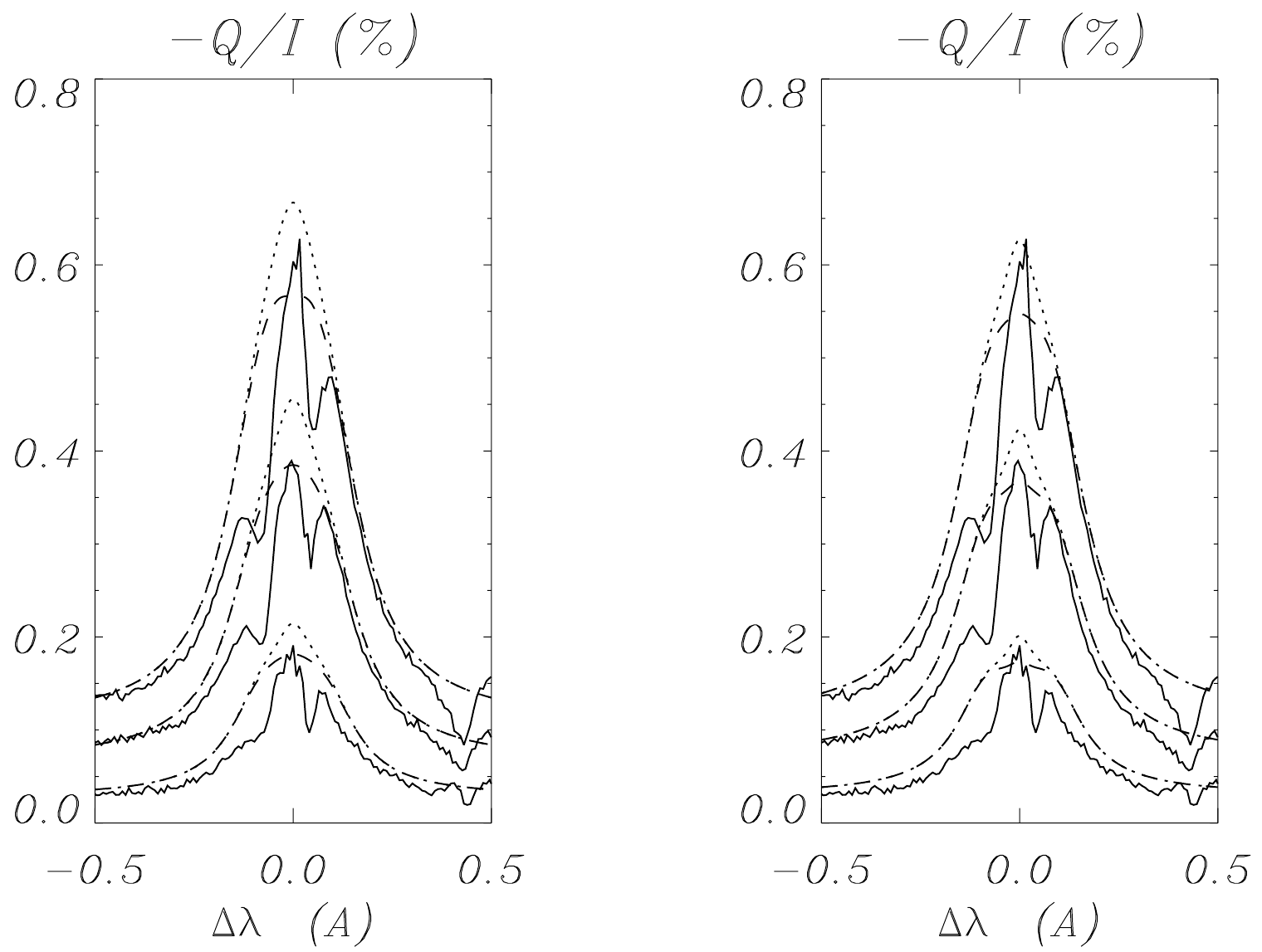}
   \caption{Hanle effect of a microturbulent magnetic field.
   Left panel:  observations versus
   non-LTE modeling for the FALC model, Right panel: observations versus
   non-LTE modeling for the FALX model.
 Full lines:
observed profiles, at 3 different distances from the solar limb,
namely, from top to bottom, 4", 10" and 40". Dashed lines:
calculated profiles for the same limb distances for B=30 G. Dotted
lines: calculated profiles for the same limb distances for B=20 G.}
\label{hanle}
    \end{figure}

Let us now turn to the Hanle effect in the
equivalent two-level atom model. In Fig. (\ref{hanle}) we compare
the polarization profiles observed at three distances from the solar
limb, namely at 4", 10" and 40", to those derived from non-LTE
radiative transfer modeling with partial frequency redistribution
and the Hanle effect, for the atmospheric models FALC and FALX. We
first remark that the observations in the line wings, which are not
sensitive to the Hanle effect are well recovered in both cases. We
recall that the central peak, due to the even isotopes, is sensitive
to the Hanle effect of a microturbulent magnetic field, whereas the
two secondary peaks, due to the odd isotopes, are not (Belluzzi {\it
et al.} 2007). We can thus focus on the interpretation of the
polarization observed in the central peak, which is modeled here. We
see that it is quite well recovered when we introduce the
depolarizing Hanle effect of a turbulent magnetic field between 20
Gauss and 30 Gauss. The differences between FALC and FALX results
are not meaningful with regard to the accuracy of the polarization
measurements.

We now take into account the effect of neglecting alignment transfer
with the metastable $^2D_{3/2,5/2}$ levels. Table 1 gives the
correction factor $f_c$ that we have to apply to the value of the
turbulent magnetic field strength derived from our equivalent
two-level model. It is obtained as explained in Derouich (2008) by
solving the statistical equilibrium equations for the density matrix
elements of the 5-level model of Ba\, {\sc ii} and of the two-level
model, in the presence of depolarizing collision, a radiation field
and the Hanle effect and comparing the alignment of the $^2P_{3/2}$
level in both models. The line profiles observed close to the solar
limb, are formed at altitudes $z \ge$ 900 km, where the hydrogen
density is lower than $5\,10^{13}$ cm$^{-3}$ and the correction
factor is between 0.60 and 0.51.  Let us stress that the correction
factor depends in a non-linear way on the magnetic field strength
implemented in the density matrix statistical rates; here it was
computed for a magnetic field of the order of 15 G. Applying this
correction to our previous estimate we find that the turbulent
magnetic field strength is between 10 G and 18 G.

\section{Conclusion}
The chromosphere is a highly inhomogeneous medium where temperature,
velocity and magnetic fields have complex structures that are still
poorly investigated. As the magnetic strength decreases with height
in the solar atmosphere, the Hanle effect in resonance lines offers
a valuable complement to Zeeman diagnostics. We have tested the
sensitivity of the linear resonance polarization of the BaII 455.4
nm line to partial frequency redistribution (PFR), temperature
variations in the atmospheric model, elastic collisions  and weak
unresolved magnetic fields. These investigations show that the line
polarization is strongly affected by PFR, but that it is weakly
sensitive to temperature variations. Recent studies on the
depolarizing effects of elastic collisions with hydrogen atoms by
Derouich (2008) showed that the line polarization is affected by
alignment transfer with the metastable $^2D$ levels which are not
included in the equivalent two-level atom model that we use for
modeling the non-LTE polarized line formation. We propose to take
this depolarizing mechanism into account by introducing a correction
factor on the turbulent magnetic field strength obtained with the
equivalent two-level atom model. We estimate this correction factor
by solving the statistical equilibrium equations for the density
matrix of the 5-levels atomic model including collisions, radiation
and magnetic field effects, with a single scattering approximation
as in Derouich (2008). This procedure is a first step toward a
complete treatment of non-LTE polarized transfer including partial
frequency redistribution and full multi-level coupling.

This paper shows that the solar Ba\,{\sc ii} D2 line is very well
suited for the diagnosis of weak magnetic fields of the order of a
few tens of Gauss in the low chromosphere, typically between 900 km
and 1300 km above the basis of the photosphere. Our observations are
well recovered with a turbulent magnetic field between 10 G and 18
G. These values are significantly lower than those which are derived
from the linear limb polarization observed in the SrI line at 406.7
nm or in molecular lines of MgH, which range between 20 G and 50 G
(see Faurobert et al 2001, Bommier et al. 2005, Bommier et al. 2006)
. But those lines are formed in the upper photosphere, typically
between 200 km and 400 km above the basis of the photosphere, where
the turbulent magnetic field may very well be stronger than at
higher altitudes. We may consider our result as the first
quantitative indication of such a decrease of the turbulent magnetic
field strength with altitude. Further observations with a better
spatial resolution should be performed to measure simultaneously the
four Stokes parameters in the BaII line, in order to take advantage
of both Hanle and Zeeman effects to obtain a complete view of the
magnetic field structure.

%%% Text of acknowledgements runs on after this command.
\acknowledgements The observational data shown in this paper were
obtained from a campaign performed at THEMIS S.L. operated on the
island of Tenerife by CNRS-CNR in the Spanish Observatorio del Teide
of the Instituto de Astrofisica de Canarias.

%%% THE BIBLIOGRAPHY
%%%
%%% CONSULT SECTION 3 OF "INSTRUCTIONS FOR AUTHORS" FOR HOW TO USE NATBIB.
%%% AUTHORS ARE ENCOURAGED TO USE EITHER THE "THEBIBLIOGRAPY" ENVIRONMENT
%%% BY UNCOMMENTING (DELETING THE "%" SYMBOL) THE COMMANDS BELOW, OR BY
%%% USING THE BIBTEX ENVIRONMENT. TO FIND OUT WHICH IS APPLICABLE TO YOUR
%%% CONTRIBUTION, CONSULT THE VOLUME EDITORS FOR YOUR PROCEEDINGS.
%%%


\begin{thebibliography}{}

\bibitem{}Avrett, E.H. 1995 in Infrared tools for Astrophysics:
What's next?, J.R Kuhn \& M.J. Penn eds. (Singapore: World
Scientific), 303

\bibitem{}Barklem, P.S. \& O'Mara, B.J. 1998, \mnras, 300, 863

\bibitem{}Belluzzi, L., Trujillo Bueno, J. \& Landi Degl'Innocenti,
E. 2007, \aap, 666, 588

\bibitem{}Bommier, V. 1997a, \aap, 328, 706

\bibitem{}Bommier, V. 1997b, \aap, 328, 726


\bibitem[2002]{Bommier-Molodij-02}  Bommier, V., \& Molodij, G. 2002, A\&A,
381, 241

\bibitem[2002]{Bommier-Rayrole-02}  Bommier, V., \& Rayrole, J. 2002, A\&A,
381, 227

\bibitem{}Bommier, V., Derouich, M., Landi degl'Innocenti, E., Molodij,
G., Sahal-Bréchot, S. 2005, \aap, 432, 295

\bibitem{}Bommier, V., Landi Degl'Innocenti, E., Molodij, G.
2006, \aap, 458, 625


\bibitem{}Derouich, M., 2008, \aap, 481, 845



\bibitem{}Derouich, M., Sahal-Br\'echot, S. \& Barklem, P. S. 2004,
\aap, 426, 707

\bibitem{}Domke, H. \& Hubeny, I. 1988, \apj, 65, 527

\bibitem{}Faurobert-Scholl, M. 1992, \aap, 258, 521

\bibitem{}Faurobert, M., Arnaud, J., Vigneau, J. \& Frisch, H. 2001,
 \aap, 378, 627


\bibitem[1999]{Fluri-Stenflo-99}  Fluri, D., \& Stenflo, J.O., 1999 A\&A,
341, 902

\bibitem{}Fluri, D.M., Nagendra, K.N. \& Frisch, H. 2003, \aap, 303

\bibitem{}Fontenla, J. M., Avrett, E. H. \& Loeser, R. 1993, \apj, 406, 319

\bibitem{}Holtzreuter, R., Fluri, D.M. \& Stenflo, J.O. 2006, \aap,
449, L41

\bibitem{}Innes, D. E., Lagg, A. \& Solanki, S. A. (eds.) 2005,
Chromospheric and coronal magnetic fields, ESA SP596


\bibitem{}Malherbe, J. M., Moity, J., Arnaud, J. \& Roudier, Th., 2006, \aap, 462, 753

\bibitem{}Rutten, R. J. 1978, Sol. Phys., 56, 237

\bibitem{}Stenflo, J.O., 1994  Solar magnetic fields, Kluwer, Dordrecht,
p. 217

\bibitem{}Stenflo, J.O. \& Keller, C.U. 1997, \aap, 321, 927

\bibitem{}Uitenbroek, H. \& Bruls, J.H.M.J 1992, \aap, 265, 268


\end{thebibliography}
\end{document}